\documentclass[11pt,fleqn]{article}

\usepackage{amssymb,amsmath}
\usepackage{natbib}
\usepackage{vmargin,setspace,verbatim}
\usepackage{graphicx,graphics,color,tikz,tikz-network}
\usepackage{microtype}

\usepackage{libertine}
\usepackage[libertine]{newtxmath}

\definecolor{darkgreen}{rgb}{0,0.2,0}
\definecolor{darkred}{rgb}{0.3,0,0}

\usepackage[colorlinks=true, pdfstartview=FitV, linkcolor=darkgreen, citecolor=darkred, urlcolor=black]{hyperref}

\newcounter{llst}

\newcounter{llist}

\newtheorem{theorem}{Theorem}[section]

\newtheorem{definition}[theorem]{Definition}
\newtheorem{expl}[theorem]{Example}

\newtheorem{lemma}[theorem]{Lemma}

\newtheorem{dscrpt}[theorem]{Description}

\newcounter{axiomatiser}

\newcounter{myclaimcount}

\newenvironment{example}{\begin{expl} \rm}{\hfill $\blacklozenge$
\end{expl}}

\setpapersize{A4}
\setmarginsrb{80pt}{40pt}{80pt}{60pt}{12pt}{10pt}{12pt}{30pt}

\begin{document}


\title{\textbf{A Generalised $\lambda$-Core Concept \\ for Normal Form Games}}

\author{Subhadip Chakrabarti\thanks{Department of Economics, The Queen's University of Belfast, Riddel Hall, 185 Stranmillis Road, Belfast, BT9~5EE, UK. \textsf{Email: s.chakrabarti@qub.ac.uk}} \and Robert P.~Gilles\thanks{Department of Economics, The Queen's University of Belfast, Riddel Hall, 185 Stranmillis Road, Belfast, BT9~5EE, UK. \textsf{Email: r.gilles@qub.ac.uk}} \and \and Lina Mallozzi\thanks{Department of Mathematics and Applications, University of Naples Federico II, Via Claudio 21, 80125 Naples, Italy. E-mail: \textsf{mallozzi@unina.it}} }

\date{August 2024}

\maketitle

\begin{abstract}
\singlespace
\noindent
We develop a generalisation of the $\lambda $-Core solution for non-cooperative games in normal form. We show that this generalised $\lambda $-Core is non-empty for the class of separable games that admit a socially optimal Nash equilibrium. Examples are provided that indicate that non-emptiness of the generalised $\lambda $-Core cannot be expected for large classes of normal form games.

\end{abstract}

\begin{description}
\singlespace
\item[Keywords:] Core; non-cooperative game; separable game; Nash equilibrium; social optimum.

\item[JEL classification:] C71; C72
\end{description}




\section{Introduction: Cores of non-cooperative games}

This paper explores Core solutions for normal form non-cooperative games, focusing on the development of well-defined characteristic functions derived from coalition payoffs within these games. These characteristic functions are based on various assumptions regarding the responses of non-coalition members, resulting in different coalition values. As a result, multiple distinct characteristic functions can be constructed for the class of normal form games. This multiplicity leads to various interpretations of what constitutes the Core of a normal form non-cooperative game.

We first explore the different construction methods to build characteristic functions describing these related cooperative game-theoretic representations of normal form games. \citet{Aumann1959}  seminally proposed the $\alpha $- and $\beta $-characteristic functions based on the assumption of min-max behaviour by players outside a considered coalition. Aumann introduced the $\alpha$-characteristic function for normal form games by positing that each coalition within such games strives to secure the highest possible payoff, regardless of the strategies adopted by players outside the coalition. This concept can be likened to a sequential decision-making process: the coalition under consideration acts first, selecting a strategy that maximises its own payoff. It then anticipates that non-member coalitions might subsequently choose strategies that minimise this maximum payoff, leading to a formulation of the characteristic function in terms of maximin strategies.

A second framework to determine the collective wealth that can be assigned to a coalition in a normal form game was also considered by \citet{Aumann1959}. The $\beta$-characteristic function determines the highest achievable payoff for a coalition, independent of the strategies adopted by outsiders. Under a framework of sequential decision-making, it is assumed that non-member coalitions act initially, selecting strategies that minimise the payoff for the considered coalition. Conversely, the considered coalition subsequently chooses strategies to maximise its collective payoff. Consequently, the resulting $\beta$-characteristic function is based on a minimax formulation. \citet{Zhao1999Beta} provided existence results of $\beta $-Core solutions for oligopolies.

The notion of the $\gamma $-characteristic function of a normal form game, developed by \citet{Chander1997}, is a departure from the framework set of the $\alpha $- and $\beta $-conceptions, based on the assumption that non-members of a coalition play individual best replies to the chosen collective strategy of the coalition and all other non-members. \citet{Helm2001} generalised the non-emptiness of the $\gamma$-Core to a larger class of games with externalities. \citet{Lardon2012,Lardon2020} applied the $\gamma $-Core solution concept to Cournot oligopoly games, while \citet{Lardon2019}  investigated the $\gamma $-Core for interval oligopolistic cooperative games, i.e., games where each coalition is assigned an interval of possible worths. \citet{Stamatopoulos2016} looked at the $\gamma $-Core for the particular class of aggregative games.

The $\delta $-characteristic function of a normal form game, developed by \citet{Currarini2015} based on the ideas of \citet{HartKurz1983}, is closely related to the $\gamma $ formulation. Instead of breaking up in singleton players, the complement of the coalition under consideration remains in tact and seeks to optimise its collective payoff.  Non-emptiness of the $\delta$-Core was investigated by \citet{ReddyZaccour2016} for the class of games exhibiting multilateral externalities as considered in \citet{Chander1997}.

\citet{Currarini2003,Currarini2015} also presented a refinement of the $\gamma $-formulation introduced as the $\lambda $-characteristic function. In particular, the $\lambda $-characteristic function is founded on the idea that a coalition should have a first mover advantage. Hence, a coalition is assumed to be a Stackelberg leader in relation to its complement. This implies that the $\lambda $-characteristic function is founded on a subgame perfect reasoning. \citet{Currarini2003,Currarini2015} consider the $\lambda $-characteristic function only under assumptions that impose that the reaction of the followers exists and is unique. 

In this study, we extend the $\lambda$-characteristic function to a broader class of games. Specifically, we eliminate the assumption that the optimal actions of non-coalition members yield a unique best response. By discarding this uniqueness requirement, we offer a generalisation of the $\lambda$-Core, identifying a class of normal form games that accommodate such generalised $\lambda$-Core solutions. This class encompasses separable games that possess a socially optimal Nash equilibrium. We demonstrate that, under specific conditions, this equilibrium produces a core allocation for the generalised $\lambda$-characteristic function.

Furthermore, we demonstrate the challenge in generalising the existence theorem to encompass a wider range of games. We develop some counterexamples illustrating that failure to satisfy certain conditions outlined in the existence theorem can lead to empty generalised $\lambda$-Cores.

\section{Some preliminaries}

Consider a given finite set of players $N=\{1,\ldots ,n\}$. Every player $i\in N$ is assigned a \emph{strategy set} denoted by $A_{i} \subseteq \mathbb{R}^{k_{i}}$, being a compact subset of some Euclidean space. The \emph{payoff function} of player $i\in N$ is a continuous map $u_{i}\colon A\rightarrow \mathbb{R}$ that assigns to every strategy profile $a\in A=A_{1}\times A_{2}\times \cdots \times A_{n}$ a payoff $u_{i}(a)$. We denote by $u=(u_{1},\ldots ,u_{n})\colon A\to \mathbb{R}^{n}$ the tuple of individual payoff functions. The pair $\Gamma =(A,u)$ is a \emph{normal form game} on player set $N$. 

We refer to a strategy profile $\hat{a}\in A$ as a \emph{social optimum} in the normal form game $\Gamma $ if $\sum_{i\in N}u_{i}(\hat{a})\geqslant \sum_{i\in N}u_{i}(a)$ for all $a\in A$ \citep{Chinchuluun2008}.

For every player $i \in N$ and strategy profile $a \in A$ we denote by $a_{-i} \in \prod_{j \neq i} A_j$ the strategy profile except player $i$'s given by $a_{-i} = (a_1, \ldots , a_{i-1}, a_{i+1}, \ldots ,a_n)$. A strategy profile $a^* \in A$ is a \emph{Nash equilibrium} \citep{Nash1950} of the game $\Gamma$ if for every $i \in N \colon u_i (a^*) \geqslant u_i (b_i, a^*_{-i} )$ for any $b_i \in A_i$.

\paragraph{Best response structures}
An alternative definition of the Nash equilibrium concept can be given through the best response structure in a game $\Gamma = (A,u)$. A \emph{best response} of player $i$ with regard to $a_{-i}\in A_{-i}$ is a strategy $\hat{a}_{i}\in A_{i}$ such that $u_{i}(\hat{a}_{i},a_{-i})\geqslant u_{i}(b_{i},a_{-i})$ for all $b_{i}\in A_{i}$. The resulting \emph{best response correspondence} for player $i$ is a map $B_{i}\colon A_{-i} \rightarrow \mathcal{P}(A_{i})$ given by 
\begin{equation}
	B_{i}(a_{-i})=\arg \max_{a_{i}\in A_{i}}\,u_{i}(a_{i},a_{-i})
\end{equation}%
for all $a_{-i}\in A_{-i}$.\footnote{For any set $X$, we denote by $\mathcal{P}(X)=\{Y|Y\subseteq X\}$ the collection of all subsets of $X$. It is called the power set of $X$.} We refer to $B=\prod_{i=1}^{n}B_{i} \colon A\rightarrow A$ as the \emph{best response correspondence} of $\Gamma $.

\citet{Nash1950} showed that a strategy profile $a^* \in A$ is a Nash equilibrium of $\Gamma$ if and only if $a^*$ is a fixed point of the best response correspondence $B$, i.e., $a^* \in B (a^* )$.

More generally, for any coalition of players $S \subseteq N$ we define $A_{S}=\prod_{i\in S} A_{i}$. We can now write any strategy tuple $a\in A$ as $a=(a_{S},a_{N\setminus S})\in A_{S}\times A_{N\setminus S}$. Furthermore, we let $a_{-S}\in A_{-S}=A_{N\setminus S}$ denote the collective strategy of the complement of $S$ in $\Gamma $. A \emph{best response} of coalition $S \subseteq N$ with regard to $a_{-S}\in A_{-S}$ is a strategy $\hat{a}_{S}\in A_{S}$ such that $\sum_{i\in S}u_{i}(\hat{a}_{S},a_{-S})\geqslant \sum_{i\in S}u_{i}(b_{S},a_{-S})$ for all $b_{S}\in A_{S}$. The resulting \emph{best response correspondence} for coalition $S$ is the map $B_{S}\colon A_{-S}\rightarrow \mathcal{P}(A_{S})$ given by 
\begin{equation}
	B_{S}(a_{-S})=\arg \max_{a_{S}\in A_{S}}\,\sum_{i\in S}u_{i}(a_{S},a_{-S}).
\end{equation}
Now, a strategy profile $a^{\ast }\in A$ is a \emph{strong Nash equilibrium} \citep{Aumann1959} if for all coalitions $S \subseteq N$ it holds that $a_{S}^{\ast}\in B_{S}(a_{-S}^{\ast })$.

\paragraph{Characteristic functions}

Throughout this paper we investigate normal form games from the perspective of coalitional ability to achieve collective payoffs. For that purpose we introduce characteristic functions that quantify these coalitional abilities. Formally, a \emph{characteristic function} on player set $N$ is a function $v\colon 2^{N}\rightarrow \mathbb{R}$ that assigns to every coalition $S\subseteq N$ a collective payoff $v(S)$ with $v(\varnothing )=0$.

An \emph{allocation} for characteristic function $v$ is defined as a vector $x\in \mathbb{R}^{N}$ such that $\sum_{i\in N}x_{i}=v(N)$. The halfspace of all allocations for $v$ is defined by $\mathbb{A}(v)\subset \mathbb{R}^{N}$. For coalition $S \subseteq N$ and allocation $x \in \mathbb A (v)$ we define $x(S) = \sum_{i \in S} x_i$.  An allocation $x\in  \mathbb{A}(v)$ is an \emph{imputation} for $v$ if for all players $i\in N\colon x_{i}\geqslant v(\{i\})$.

The \emph{Core} for the characteristic function $v$ \citep{Gillies1959} is the collection of all allocations that pay every coalition $S$ at least their assigned worth $v(S)$, i.e., the Core is defined by
\begin{equation}
	C (v) = \left\{ x \in \mathbb A(v) \mid x(S) \geqslant v(S) \mbox{ for every } S \subseteq N \right\} .
\end{equation}
The core of a coalition game assumes the collective payoff of a coalition is fixed and independent of the specific method used to determine it. It focuses on whether the coalition can distribute this payoff among its members in a way that prevents any subgroup from being better off by forming a separate coalition.

In contrast, normal form games explore strategic interactions among coalitions and non-members to determine collective payoffs, leading to diverse procedural considerations represented by different characteristic functions. This complexity underscores the variety of core concepts that emerge based on strategic behaviours within normal form game settings.

We provide an overview of the main characteristic functions for normal form games that were developed in the literature. In particular, for every coalition $S \subset N$ the  following table provides such a survey:
\begin{center}
\begin{tabular}{|l|l|} \hline
 {\bf Solution concept }  &    {\bf  Characteristic function}     \\ \hline 
 \ \ \textbf{\textcolor{blue}{$\alpha$-core}} \citep{Aumann1959}   &   {$ v^{\alpha} (S) = \displaystyle  \max_{a_S \in A_S} \min_{b_{-S} \in A_{-S}} \sum_{i \in S} u_i (a_S, b_{-S})$   }  
    \\ \hline 
  \ \ \textbf{\textcolor{blue}{$\beta$-core}} \citep{Aumann1959}   &   {$  v^{\beta} (S) =  \displaystyle  \min_{b_{-S} \in A_{-S}} \max_{a_S \in A_S} \sum_{i \in S} u_i (a_S, b_{-S})$   }  
    \\ \hline 
  \ \ \textbf{\textcolor{blue}{$\gamma$-core}} \citep{Chander1997}  &   {$   v^\gamma (S) = \sum_{i \in S} u_i ( \hat{a}^S )$, $\hat{a}^S \in \displaystyle \arg \max_{a^\star \in \mathsf{E}^\gamma (S)} \, \sum_{i \in S} u_i (a^\star ) $   }  
   \\ \hline 
  \ \ \textbf{\textcolor{blue}{$\delta$-core}} \citep{Currarini2015}	 &   {$   v^\delta (S) = \sum_{i \in S} u_i ( \tilde{a}^S ), \tilde{a}^S \in \displaystyle  \arg \max_{a^\star \in \mathsf{E}^\delta (S)} \, \sum_{i \in S} u_i (a^\star )$   }  
   	\\ \hline 
\end{tabular}
\end{center}
while  $v^{\alpha} (N) =  v^{\beta} (N) = v^\gamma (N)= v^\delta (N) = \displaystyle \max_{a \in A} \, \sum_{i \in N} u_i(a)$,  where  
\[
\mathsf{E}^\gamma (S) = \left\{ a^\star \in A \, \left| \, 
\begin{array}{c}
	\mbox{For all } a_S \in A_S \colon \sum_{i \in S} u_i (a^\star ) \geqslant \sum_{i \in S} u_i (a_S , a^\star_{-S} ) , \\
	\mbox{and for every } j \in N \setminus S, \ a_j \in A_j \colon u_j (a^\star ) \geqslant u_j (a_j , a^\star_{-j} )
\end{array}
\right. \right\}
\]
and
\[
\mathsf{E}^\delta (S) = \left\{ a^\star \in A \, \left| \, 
\begin{array}{c}
	\mbox{For all } a_S \in A_S \colon \sum_{i \in S} u_i (a^\star ) \geqslant \sum_{i \in S} u_i (a_S , a^\star_{-S} ) , \\
	\mbox{and for every } a_{-S} \in A_{-S} \colon \sum_{j \in N \setminus S} \, u_j (a^\star ) \geqslant \sum_{j \in N \setminus S} \, u_j (a^\star_S , a_{-S} )
\end{array}
\right. \right\}.
\]

\section{The $\lambda$-characteristic function and its generalisation}

The term $\lambda $-characteristic function was coined by \citet{Currarini2015}\footnote{This conception was seminally introduced by the same authors in \citet{Currarini2003}.} The $\lambda $-Core of a normal form game is founded on a similar principles as the $\gamma $-Core, except that the coalition under consideration is assumed to have a leadership position in the execution of chosen strategies and, therefore, has a first-mover advantage over the players that are not member of that coalition. In that respect, it is also natural to refer to the $\lambda $-Core as the $\gamma $-Core with a Stackelberg leader \citep{Stamatopoulos2020}.

The $\lambda $-characteristic function was developed for a specific subclass of normal form games satisfying the \emph{Strong Reduction Property}. In these games, it is assumed that the sequential (Stackelberg) structure is trivial, since decisions by any coalition are assumed to result in a unique optimal and stable choice for the players outside the coalition. Hence, there exists a unique Nash equilibrium if the strategy profile for players in any given coalition is fixed. This is a rather strong hypothesis.

We seek to weaken the Strong Reduction Property and to consider situations where there are possibly multiple Nash equilibria for players outside a given coalition with a fixed strategic profile.

\subsection{The Strong Reduction property}

Let $\Gamma =(A,u)$ be some normal form game on player set $N$. Considering any non-empty coalition $\varnothing \neq S \subseteq N$, if the coalition commits to the coalitional strategy $\bar a_{S}\in A_{S}$, then this is equivalent to coalition $S$ leaving the game by implementing the collective strategy $\bar a_{S} $, resulting in a \emph{reduced game} based on $\Gamma $ characterised by (1) the reduced player set $N\setminus S$; (2) the reduced strategy profile set $A_{-S}$, and; (3) the modified payoff structure $\bar w^{S, \bar a_S}\colon A_{-S}\rightarrow \mathbb{R}^{N}$ defined by $\bar w_{j}^{S}(b_{-S})=u_{j}(\bar a_{S},b_{-S})$ for every player $j\in N\setminus S$ and strategy profile $b_{-S}\in A_{-S}$. 

The set of all Nash equilibria of the reduced game $(N\setminus S,A_{-S},\bar w^{S, \bar a_S})$ is now denoted by $\mathsf{E}(N\setminus S,A_{-S}, \bar w^{S, \bar a_S })\subseteq A_{-S}$. Note that this set can be empty. The Strong Reduction property not only excludes non-emptiness, but also imposes that there exists a \emph{unique} Nash equilibrium in each of these subgames.
\begin{definition} \textbf{\emph{(Strong Reduction Property)}} \\
	A normal form game $\Gamma =(A,u)$ has the \textbf{Strong Reduction Property} if for every non-empty coalition $\varnothing \neq S \subset N$ and every coalitional strategy $\bar a_{S}\in A_{S}$ for $S$ it holds that $\mathsf{E} (N\setminus S,A_{-S}, \bar w^{S,\bar a_S})\subseteq A_{-S}$ is a singleton, i.e., there exists a unique Nash equilibrium in the $(S,\bar a_{S})$-reduced game.
\end{definition}
Next, assume that $\Gamma =(A,u)$ is a normal form game that has the Strong Reduction Property. For every non-empty coalition $\varnothing \neq S \subseteq N$ and coalitional strategy $\bar a_{S}\in A_{S}$, let $\mathsf{E}(N\setminus S,A_{-S}, \bar w^{S,\bar a_S})=\left\{ \bar{b}_{-S}(\bar a_{S})\right\} $. Then the reduced payoff function for coalition $S$ is given by $\bar w \colon A_{S}\rightarrow \mathbb{R}$ defined by 
\begin{equation}
	\bar w ( \bar a_{S}) = \sum_{i\in S}u_{i}\left( \, \bar a_{S},\bar{b}_{-S}( \bar a_{S})\,\right) .
\end{equation}
The $\lambda $-\textbf{characteristic function} for the game $\Gamma $ is defined by $v^{\lambda }(\varnothing )=0$, for every non-empty coalition $\varnothing \neq S \subset N$ by 
\begin{equation}
	v^{\lambda }(S)=\max_{\bar a_{S}\in A_{S}} \bar w ( \bar a_{S}),
\end{equation}%
and 
\begin{equation}
	v^{\lambda }(N)=\max_{a\in A}\sum_{i\in N}u_{i}(a).
\end{equation}%
The $\lambda $-\textbf{Core} of $\Gamma $ is now given by $\mathcal{C}^{\lambda}(\Gamma )=\mathcal{C}(v^{\lambda })$.

We conclude the introduction of the notion of the $\lambda $-core of a normal form game by stating the comparative result from \citet{Currarini2003} and \citet{Chander2010}.
\begin{lemma}
	Let $\Gamma $ be a normal form game that satisfies the Strong Reduction Property. Then the following holds:
\begin{equation}
	\varnothing \neq \mathcal{C}^{\lambda }(\Gamma )\subseteq \mathcal{C}^{\gamma }(\Gamma )\subseteq \mathcal{C}^{\beta }(\Gamma )\subseteq \mathcal{C}^{\alpha}(\Gamma ).
\end{equation}
\end{lemma}
We emphasise that the Strong Reduction Property is a requirement that imposes severe restrictions on the class of games for which the $\lambda$-Core is properly defined.

\subsection{A generalised $\lambda$-Core of a normal form game}

We aim to extend the definition of the notion of the $\lambda $-core beyond the very limited realm of games that satisfy the Strong Reduction Property. For that purpose we generalise the definition of the $\lambda $-characteristic function to arbitrary normal form games.

Let $\Gamma =(A,u)$ be some normal form game on player set $N=\{1,\ldots ,n\}$. Again, let for some $\varnothing \neq S \subset N$ and strategy $\bar a_S \in A_S$, $\mathsf{E}(N\setminus S,A_{-S}, \bar w^{S, \bar a_S})\subseteq A_{-S}$ denote the set of Nash equilibria in the reduced game $(N\setminus S, A_{-S}, \bar w^{S, \bar a_S})$ as introduced in the discussion of the $\lambda $-characteristic function. Now, we define the \emph{generalised $\lambda $-characteristic function} as $\bar{v}^{\lambda }\colon 2^{N}\rightarrow \mathbb{R}$ by $\bar{v}^{\lambda }(\varnothing )=0$ and for every $\varnothing \neq S \subset N \colon $
\begin{equation}
	\bar{v}^{\lambda }(S)=\left\{ 
	\begin{array}{ll}
		\underset{a_{S}\in A_{S}}{\max } \ \ \underset{b_{-S}\in \mathsf{E} (N\setminus S,A_{-S}, \bar w^{S, a_S})}{\max } \ \ \sum_{i\in S}u_{i}(a_{S},b_{-S}) & \text{if }\mathsf{E}(N\setminus S,A_{-S},\bar w^{S, \bar a_S})\neq \varnothing ; \\ 
		-\infty  & \text{if }\mathsf{E}(N\setminus S,A_{-S},\bar w^{S, \bar a_S})=\varnothing ;
	\end{array}%
	\right. 
\end{equation}%
and, finally,
\begin{equation}
\bar{v}^{\lambda }(N)=\max_{a\in A}\,\sum_{i\in N}u_{i}(a),
\end{equation}%
We emphasise that, obviously, we no longer impose the Strong Reduction Property. Therefore, $\mathsf{E}(N\setminus S,A_{-S}, \bar w^{S, \bar a_S})\subseteq A_{-S}$ can be empty or consist of any number of
elements.

The \emph{generalised $\lambda $-Core} of the game $\Gamma $ is now defined by
\begin{equation}
	\widehat{\mathcal{C}}^{\lambda }(\Gamma )= \{ a \in A \mid u(a) \in \mathcal{C}(\bar{v}^{\lambda }) \, \}
\end{equation}
 
\paragraph{The generalised $\lambda$-Core of separable games}

The subclass of \emph{separable} normal form games was introduced by \citet{Balder1997} and further developed and discussed by \citet{Peleg1998}. Subsequently, \citet{Milchtaich2009} considered separable congestion games with linear variable cost structures. We formalise the definition of separability as follows.
\begin{definition}
A normal form game $\Gamma =(A,u)$ on the player set $N$ is \textbf{separable} if for all pairs of players $i,j\in N$ there exist functions $h_{j}^{i}\colon A_{j}\rightarrow \mathbb{R}$ such that for every strategy profile $a\in A\colon $ 
	\begin{equation}
		u_{i}(a)=\sum_{j\in N}h_{j}^{i}(a_{j})
	\end{equation}
\end{definition}
Note that $h_{i}^{i}(a_{i})$ is a self-referential payoff in this definition of a separable game. Without proof we state the following property that follows immediately from the definition of separability. 
\begin{lemma} \label{lem:B_j}
	For any separable game $\Gamma =(A,u)$ it holds that for every player $i\in N$ and every strategy profile $a\in A \colon$
\begin{equation*}
	B_{i}(a_{-i})=\arg \max_{b_{i}\in A_{i}}u_{i}(b_{i},a_{-i})=\arg \max_{b_{i}\in A_{i}}\left[ \,h_{i}^{i}(b_{j})+\sum_{j\neq i}h_{j}^{i}(a_{j})\,\right] =\arg \max_{b_{i}\in A_{i}}h_{i}^{i}(b_{i}).
\end{equation*}
This implies that $B_i (a_{-i}) = B_i \subseteq A_i$ for any $a_{-i} \in A_{-i}$.
\end{lemma}
Our main result shows that for separable games that admit a socially optimal Nash equilibrium, the generalised $\lambda $-Core is non-empty provided certain additional regularity conditions are satisfied. These regularity conditions require the maximum of additive functions to be the sum of the separate maxima.\footnote{We remind that the sum of the maximums of several functions is at least equal to the maximum of their sum over the same domain. For further discussion we refer to \citet{Chinchuluun2008}.}
\begin{theorem} \label{thm:Main} 
Let $\Gamma =(A,u)$ be a separable normal form game on player set $N$ that admits a socially optimal Nash equilibrium, i.e., there exist a Nash equilibrium $a^{\star }\in A$ such that $\sum_{i\in N}u_{i}(a^{\star })=\bar{v}^{\lambda }(N)$. Then if for all coalitions $S\subseteq N$ and players $j\in N,$
	\begin{equation} \label{1}
		\underset{a_{j}\in A_{j}}{\max }\sum_{i\in S}h_{j}^{i}(a_{j})=\sum_{i\in S}\underset{a_{j}\in A_{j}}{\max }h_{j}^{i}(a_{j}) 
	\end{equation}%
	and 
	\begin{equation} \label{2}
		\underset{a_{j}\in B_{j}}{\max }\sum_{i\in S}h_{j}^{i}(a_{j})=\sum_{i\in S}\underset{a_{j}\in B_{j}}{\max }h_{j}^{i}(a_{j}), 
	\end{equation}%
	it holds that $\widehat{\mathcal{C}}^{\lambda }(\Gamma )\neq \varnothing $.
\end{theorem}
For the proof of this theorem we refer to the next subsection. We complete our discussion of the generalised $\lambda$-Core of a separable game through a number of counter examples that show the emptiness of the generalised $\lambda $-Core if the conditions stated in Theorem \ref{thm:Main} are not satisfied. The first example considers a non-separable game in which the generalised $\lambda $-Core is indeed empty.

\begin{example} \label{ex:First} 
Let $N=\{1,2\}$ and define the normal form game $\Gamma_1 $ on $N$ by $A_{1}=A_{2}=[0,1]$ and a payoff structure with for every $a=(a_{1},a_{2})\in A_{1}\times A_{2}=[0,1]^{2}\colon u_{1}(a_{1},a_{2})=(1-a_{1}-2a_{2})a_{1}$ and $u_{2}(a_{1},a_{2})=(1-2a_{1}-a_{2})a_{2}$. \\
In this case, we derive that the best response correspondences $B_{1}$ and $B_{2}$ are actually continuous functions given by 
\begin{equation*}
	B_{1}(a_{2})=\left\{ 
	\begin{array}{ll}
		\tfrac{1}{2}-a_{2} & \text{if }0\leqslant a_{2}<\tfrac{1}{2}; \\ 
		0 & \text{if }\tfrac{1}{2}\leqslant a_{2}\leqslant 1;%
	\end{array}%
	\right. 
	\quad \text{and} \quad
	B_{2}(a_{1})=\left\{ 
	\begin{array}{ll}
		\tfrac{1}{2}-a_{1} & \text{ if }0\leqslant a_{1}<\tfrac{1}{2}; \\ 
		0 & \text{if }\tfrac{1}{2}\leqslant a_{1}\leqslant 1.%
	\end{array}%
	\right.
\end{equation*}%
From this we conclude that 
\begin{align*}
	\bar{v}^{\lambda }(\{1\})& =\max_{a_{1}\in A_{1}}\max_{a_{2}\in B_{2}(a_{1})}u_{1}(a_{1},a_{2})=\tfrac{1}{4} \\
	\bar{v}^{\lambda }(\{2\})& =\max_{a_{2}\in A_{2}}\max_{a_{1}\in B_{1}(a_{2})}u_{2}(a_{1},a_{2})=\tfrac{1}{4} \\
	\bar{v}^{\lambda }(N)& =\max_{(a_{1},a_{2})\in [0,1]^{2}}(a_{1}+a_{2})-(a_{1}^{2}+a_{2}^{2})-4a_{1}a_{2}=\tfrac{1}{4}
\end{align*}%
It is easy to check that $\widehat{\mathcal{C}}^{\lambda } (\Gamma_1)=\varnothing $.
\end{example}
The next example shows that the condition that the identified Nash equilibrium is also a social optimum, is critical. Again we use a two-player game to show this.
\begin{example} \label{ex:second}
Let $N=\{1,2\}$. We define the normal form game $\Gamma_2 $ on $N$ by $A_{1}=A_{2}=[0,1]$ and a payoff structure with for every $a=(a_{1},a_{2})\in A_{1}\times A_{2}=[0,1]^{2}\colon u_{1}(a_{1},a_{2})=a_{1}$ and $u_{2}(a_{1},a_{2})=-a_{1}^{2}-a_{2}$. \\
It is easy to establish that $B_{1}$ and $B_{2}$ are constant functions with $B_{1}(a_{2})=1$ and $B_{2}(a_{1})=0$ for all $(a_{1},a_{2})\in [0,1]^{2}$. The unique Nash equilibrium is, therefore, $(1,0)$, which is different from the unique social optimum, determined as $\left( \tfrac{1}{2} ,0\right) $. \\
Next, it can easily be determined---using the formulations given in Example \ref{ex:First}---that $\bar{v}^{\lambda }(\{1\})=1$, $\bar{v}^{\lambda}(\{2\})=-1$, and $\bar{v}^{\lambda }(\{1,2\})=\max_{(a_{1},a_{2})\in [0,1]^{2}}a_{1}-a_{1}^{2}-a_{2}=\tfrac{1}{4}$. This leads to the conclusion that $\widehat{\mathcal{C}}^{\lambda } (\Gamma_2) =\varnothing $.
\end{example}

\paragraph{Application: A status game}

\citet{Akerlof1997} and \citet{LeBretonWeber2011}  considered a status model where strategic choices of all players represent a one-dimensional interval and an individual utility depends on a comparison between her own status (the individual's behaviour)  and the status of all others within the society. We discuss a specification of this status model that admits a unique generalised $\lambda$-Core solution. 

We describe a simple status game $\Gamma_3$ as follows. Each players $i \in N$ selects a status-inducing action $a_i \in A_i = [0,1]$. Hence,  $a=(a_1, \ldots ,a_n)\in  A_1\times \cdots \times A_n = [0,1]^n$. Each player experiences a disutility $d(a_i - \bar a_{-i})$ based on the difference between her choice $a_i$ and the average of the choices of everyone else $\bar a_{-i} = \tfrac{1}{n-1} \sum_{j \neq i} a_j \in [0,1]$---being an expression of the status of the other players in the society. Here, $d \colon [-1,1] \to \mathbb R$ is assumed to be some one-dimensional disutility function. If $f_i(a_i)$ is the intrinsic value of player $i$'s action, the net payoffs are now given by $u_i(a)= f_i(a_i) - d (a_i - \bar a_{-i} )$. This defines the status game $\Gamma_3 = (\, [0,1]^n,u)$.

Assuming that $d$ is the identity function and intrinsic benefits are quadratic, in the sense that $f_i(a_i)= a_i^2$ for $i \in N$, we arrive at $u_i (a) = a^2_i - a_i + \bar a_{-i}$. For these specifications, the status game $\Gamma_3$ is separable and it is easy to establish that $B_i (a_{-i}) = B := \{ 0,1 \}$ for any $a \in A$. There are $2^n$ Nash equilibria, corresponding to the vertices of the unit hypercube in $\mathbb R^n$. Moreover, the Nash equilibrium $a^\star = (1, \ldots ,1)$ is the unique social optimum in this game.

We conclude therefore that $\bar v^\lambda (S) = |S|$ and that, by Theorem \ref{thm:Main}, $\widehat{\mathcal C}^\lambda (\Gamma_3) = \{ a^\star \}$.

\subsection{Proof of Theorem \ref{thm:Main}}

Let $\Gamma =(A,u)$ be, as postulated, a separable normal form game on player set $N$, satisfying (\ref{1}) and (\ref{2}) that admits a socially optimal Nash equilibrium $a^{\star }\in A$. Throughout we denote for every $i \in N$ the constant set of best responses by $B_i \subseteq A_i$ (Lemma \ref{lem:B_j}).
\\
Now, by assumptions (\ref{1}) and (\ref{2}),
\begin{align*}
	\bar{v}^{\lambda }(N)& =\max_{a\in A}\sum_{i\in N}u_{i}(a)=\max_{a\in A}\sum_{i=1}^{n}\sum_{j=1}^{n}h_{j}^{i}(a_{j}) =\max_{a\in A}\sum_{j=1}^{n}\sum_{i=1}^{n}h_{j}^{i}(a_{j}) \\
	& =\sum_{j=1}^{n}\left[ \,\max_{a_{j}\in A_{j}}\sum_{i=1}^{n}h_{j}^{i}(a_{j})\,\right] =\sum_{j=1}^{n}\sum_{i=1}^{n} \max_{a_{j}\in A_{j}}h_{j}^{i}(a_{j}).
\end{align*}
Given that $a^{\star }$ is a social optimum, it follows that 
\begin{equation*}
	\bar{v}^{\lambda }(N)=\sum_{j=1}^{n}\sum_{i=1}^{n}h_{j}^{i}(a_{j}^{\star }).
\end{equation*}
Furthermore, by $a^{\star }$ being a Nash equilibrium of $\Gamma $, it follows that $a_{j}^{\star }\in B_{j}$ for all $j\in N$.
\\
Next, take an arbitrary coalition $\varnothing \neq S \subset N$ and let $\bar a_S \in A_S$. We investigate the maximisation of the collective payoff of $S$ over $\mathsf{E}(N\setminus S,A_{-S}, \bar w^{S, \bar a_S})$. Now for any $b_{-S} \in A_{-S}$, player $i\in N\setminus S$ is maximising $u_{i}(\bar a_S , b_{-S})=\sum_{j\in N}h_{j}^{i}(\bar a_S , b_{-S})$ with respect to $b_{i}\in A_{i}$ and therefore will choose $b_{i}\in B_{i}$. Therefore, using the auxiliary notation $B(-S)=\prod_{j\in N\setminus S}B_{j}$,
\begin{align*}
	\max_{b_{-S}\in \mathsf{E}(N\setminus S,A_{-S},\bar w^{S, \bar a_S})}\sum_{i\in S}u_{i}(\bar a_{S},b_{-S}) =& \max_{b_{-S}\in \mathsf{E}(N\setminus S,A_{-S},\bar w^{S, \bar a_S})}\sum_{i\in S} \left[ \sum_{j\in S}h_{j}^{i}(\bar a_{j}) + \sum_{h \in N \setminus S}h_{h}^{i}(b_{h}) \right] \\
	= &\max_{b_{j}\in B_{j} \colon j\in N\setminus S}\sum_{i\in S} \left[ \sum_{j\in S}h_{j}^{i}(\bar a_{j}) + \sum_{h \in N \setminus S}h_{h}^{i}(b_{h}) \right] \\
	=&\max_{b_{-S}\in B(-S)}\sum_{i\in S} \left[ \sum_{j\in S}h_{j}^{i}(\bar a_{j}) + \sum_{h \in N \setminus S}h_{h}^{i}(b_{h}) \right]
\end{align*}
Using the above, we first show that $\bar v^{\lambda}$ is partitionally superadditive.  First, from our assumptions (\ref{1}) and (\ref{2}) it follows that
\begin{align*}
	\bar{v}^{\lambda }(S) =&\max_{a_{S}\in A_{S}}\ \max_{b_{-S}\in \mathsf{E} (N\setminus S,A_{-S},\bar w^{S, a_S})}\sum_{i\in S}u_{i}(a_{S},b_{-S}) \\[1ex]
	= & \max_{a_{S}\in A_{S}}\ \max_{b_{-S}\in B(-S)}\sum_{i\in S} \left[ \sum_{j\in S}h_{j}^{i}(a_{j}) + \sum_{h \in N \setminus S}h_{h}^{i}(b_{h}) \right] \\[1ex]
	=&\max_{a_{S}\in A_{S}}\ \sum_{j\in S}\ \sum_{i\in S}h_{j}^{i}(a_{j})+\max_{b_{-S}\in B(-S)}\sum_{j\in N\setminus S}\ \sum_{i\in S}h_{j}^{i}(b_{j}) \\[1ex]
	=&\sum_{j\in S}\,\max_{a_{j}\in A_{j}}\sum_{i\in S}h_{j}^{i}(a_{j})+\sum_{j\in N\setminus S}\ \max_{b_{j}\in B_{j}}\sum_{i\in S}h_{j}^{i}(b_{j}) \\[1ex]
	=&\sum_{j\in S}\ \sum_{i\in S}\max_{a_{j}\in A_{j}}h_{j}^{i}(a_{j})+\sum_{j\in N\setminus S}\ \sum_{i\in S}\max_{b_{j}\in B_{j}}h_{j}^{i}(b_{j})
\end{align*}%
Likewise, we derive that 
\begin{equation*}
	\bar{v}^{\lambda }(N\setminus S)=\sum_{j\in N\setminus S} \ \sum_{i\in N\setminus S}\left[ \,\max_{a_{j}\in A_{j}}h_{j}^{i}(a_{j})\,\right] +\sum_{j\in S}\ \sum_{i\in N\setminus S}\left[ \,\max_{b_{j}\in B_{j}}h_{j}^{i}(b_{j})\,\right]
\end{equation*}
Therefore, 
\begin{align*}
	\bar{v}^{\lambda }(S)+\bar{v}^{\lambda }(N\setminus S) = & \sum_{j\in S}\sum_{i\in S}\left[ \,\max_{a_{j}\in A_{j}}h_{j}^{i}(a_{j})\,\right] +\sum_{j\in N\setminus S}\sum_{i\in N\setminus S}\left[ \,\max_{a_{j}\in A_{j}}h_{j}^{i}(a_{j})\,\right] + \\[1ex]
	& \quad +\sum_{j\in N\setminus S}\ \sum_{i\in S}\left[ \,\max_{b_{j}\in B_{j}}h_{j}^{i}(b_{j})\,\right] +\sum_{j\in S}\ \sum_{i\in N\setminus S} \left[ \,\max_{b_{j}\in B_{j}}h_{j}^{i}(b_{j})\,\right] \\[1ex]
	\leqslant & \sum_{j\in S}\sum_{i\in S}\left[ \,\max_{a_{j}\in A_{j}}h_{j}^{i}(a_{j})\,\right] +\sum_{j\in N\setminus S}\sum_{i\in N\setminus S}\left[ \,\max_{a_{j}\in A_{j}}h_{j}^{i}(a_{j})\,\right] + \\[1ex]
	& \quad +\sum_{j\in N\setminus S}\ \sum_{i\in S}\left[ \,\max_{b_{j}\in A_{j}}h_{j}^{i}(b_{j})\,\right] +\sum_{j\in S}\ \sum_{i\in N\setminus S}\left[ \,\max_{b_{j}\in A_{j}}h_{j}^{i}(b_{j})\,\right] \\
	=&\sum_{j\in S}\sum_{i\in N}\left[ \,\max_{a_{j}\in A_{j}}h_{j}^{i}(a_{j})\, \right] +\sum_{j\in N\setminus S}\sum_{i\in N}\left[ \,\max_{a_{j}\in A_{j}}h_{j}^{i}(a_{j})\,\right] \\[1ex]
	=&\sum_{j\in N}\sum_{i\in N}\left[ \,\max_{a_{j}\in A_{j}}h_{j}^{i}(a_{j})\, \right] =\bar{v}^{\lambda }(N).
\end{align*}
From the above, we conclude that $\bar{v}^{\lambda }$ is indeed partitionally superadditive.

\bigskip\noindent
Next, we show that $\bar{v}^{\lambda }$ is also partitionally subadditive. From $a^{\star }\in A$ being a Nash equilibrium, $a_{i}^{\star }\in B_{i}$ for all players $i\in N$. This implies further that for every $i\in N\colon $ 
\begin{equation*}
	u_{i}(a^{\star })=\sum_{j\in N}h_{j}^{i}(a_{j}^{\star })\leqslant \sum_{j\in N}\left[ \,\max_{b_{j}\in B_{j}} \ h_{j}^{i}(b_{j})\,\right] .
\end{equation*}%
This leads to the conclusion that 
\begin{align*}
	\bar{v}^{\lambda }(N) = & \sum_{i\in N}u_{i}(a^{\star })\leqslant \sum_{i\in N}\sum_{j\in N}\left[ \,\max_{b_{j} \in B_{j} }h_{j}^{i}(b_{j})\,\right] \\[1ex]
	= & \sum_{i\in S}\sum_{j\in S}\left[ \,\max_{b_{j}\in B_{j}}h_{j}^{i}(b_{j})\, \right] +\sum_{i\in S}\sum_{j\in N\setminus S}\left[ \,\max_{b_{j}\in B_{j}}h_{j}^{i}(b_{j})\,\right] + \\[1ex]
	& +\sum_{i\in N\setminus S}\sum_{j\in S}\left[ \,\max_{b_{j}\in B_{j} }h_{j}^{i}(b_{j})\,\right] +\sum_{i\in N\setminus S}\sum_{j\in N\setminus S}\left[ \,\max_{b_{j}\in B_{j} }h_{j}^{i}(b_{j})\,\right] \\[1ex]
	\leqslant & \sum_{i\in S}\sum_{j\in S}\left[ \,\max_{b_{j}\in A_{j}}h_{j}^{i}(b_{j})\,\right] +\sum_{i\in S}\sum_{j\in N\setminus S}\left[ \,\max_{b_{j}\in B_{j}}h_{j}^{i}(b_{j})\,\right] + \\[1ex]
	& +\sum_{i\in N\setminus S}\sum_{j\in S}\left[ \,\max_{b_{j}\in B_{j} }h_{j}^{i}(b_{j})\,\right] +\sum_{i\in N\setminus S}\sum_{j\in N\setminus S}\left[ \,\max_{b_{j}\in A_{j}}h_{j}^{i}(b_{j})\,\right] \\
	= & \bar{v}^{\lambda }(S)+\bar{v}^{\lambda }(N\setminus S).
\end{align*}
Hence, we conclude that $\bar{v}^{\lambda }$ is partitionally additive or \emph{constant sum}: For every $S\in 2^{N}$ it holds that $\bar{v}^{\lambda }(N)=\bar{v}^{\lambda }(S)+\bar{v}^{\lambda }(N\setminus S)$. 

\medskip\noindent
We conclude the proof by showing that, since $\bar v^\lambda$ is constant-sum, $\widehat{\mathcal{C}} (\Gamma ) = \{ a \in A \mid u(a) \in \mathcal{C} (\bar{v}^{\lambda }) \, \}\neq \varnothing $. Indeed,
\[
\bar v^\lambda (S)= \max_{a_S \in A_S} \ \max_{b_{-S} \in \mathsf E (N \setminus S , A_{-S} , \bar w^{S,a_S})} \sum_{i \in S} u_i (a_S, b_{-S} ) \geqslant \max_{b_{-S} \in \mathsf E (N \setminus S , A_{-S} , \bar w^{S, a^\star_S})} \sum_{i \in S} u_i (a^\star_S, b_{-S} ).
\]
Since $ a^\star_{-S} \in \mathsf E (N \setminus S , A_{-S} , \bar w^{S, a^\star_S}) \colon \bar v^\lambda (S) \geqslant \sum_{i \in S} u_i (a^\star_S, a^\star_{-S} ) =  \sum_{i \in S} u_i (a^\star)$. Using the fact that $\bar v^\lambda$ is constant-sum, i.e., $\bar v^\lambda (N) = \bar v^\lambda (S) + \bar v^\lambda (N\setminus S)$, we conclude that
\[
\bar v^\lambda (N)= \sum_{i\in N} u_i(a^\star) = \bar v^\lambda (S) + \bar v^\lambda (N\setminus S) \geqslant  \sum_{i \in S} u_i (a^\star) + \bar v^\lambda (N\setminus S)
\]
This simplifies to $\sum_{i\in {N\setminus S}} u_i(a^\star)   \geqslant   \bar v^\lambda (N\setminus S)$. Next, by exchanging the role of $S$ and $N\setminus S$, we also arrive at $\sum_{i\in {S}} u_i(a^\star)   \geqslant   \bar v^\lambda ( S)$.
\\
Therefore, $\sum_{i\in {S}} u_i(a^\star)   =  \bar v^\lambda ( S)$, and in particular  $u_i(a^\star)  =  \bar v^\lambda (i)$ for any $i\in N$. Hence, $\bar v^\lambda (N)= \sum_{i\in N} \bar v^\lambda (i)$, implying that $(\bar v^\lambda (1), \ldots , \bar v^\lambda (n) \, ) = (u_1(a^\star),....,u_n(a^\star))\in \mathcal C ( \bar v^\lambda )$, leading to the desired conclusion that $a^\star \in \widehat{\mathcal C}^\lambda ( \Gamma ) \neq \varnothing$.

\section{Some concluding remarks}

The question whether the generalised $\lambda $-Core of a normal form game is non-empty has been affirmatively answered for separable games that admit socially optimal Nash equilibria provided conditions (\ref{1}) and (\ref{2}) are satisfied. This is a rather restrictive class of games. This research can possibly be extended to more broad classes of games, but general existence theorems are hard to establish.

Here, we explore insights for the more broad class of \emph{additively separable} games introduced by \citet{Mishra2018}.
\begin{definition}
A normal form game $\Gamma =(A,u)$ on the player set $N$ is \textbf{additively separable} if for every player $i\in N$ there exists a function $s_{i}\colon A_{i}\rightarrow \mathbb{R}$ such that $\sum_{i\in N}u_{i}(a)=\sum_{i\in N}s_{i}(a_{i})$ for every strategy profile $a\in A$.
\end{definition}
We conjecture that for the class of additively separable games the resulting generalised $\lambda $-Core is non-empty subject to certain regularity conditions. In particular, we expect that the generalised $\lambda$-Core is non-empty if the conditions of Theorem \ref{thm:Main} for this larger class of additively separable games hold. The next example shows this for a two-player game.
\begin{example}
Let $N=\{1,2\}$. We define the normal form game $\Gamma $ on $N$ by $A_{1}=A_{2}=[0,1]$ and a payoff structure such that for every $a=(a_{1},a_{2})\in A_{1}\times A_{2}=[0,1]^{2}\colon u_{1}(a_{1},a_{2})=\left( \tfrac{1}{2}-a_{2}\right) a_{1}$ and $u_{2}(a_{1},a_{2})=a_{1}a_{2}$. \\
In this case we derive that for every $(a_{1},a_{2})\in \lbrack 0,1]^{2}\colon $ 
\begin{equation*}
B_{1}(a_{2})=\left\{ 
\begin{array}{ll}
\{1\}\smallskip  & \text{if }0\leqslant a_{2}<\tfrac{1}{2} \\ 
\lbrack 0,1]\smallskip  & \text{if }a_{2}=\tfrac{1}{2} \\ 
\{0\} & \text{if }\tfrac{1}{2}<a_{2}\leqslant 1
\end{array}
\right. \quad \text{and} \quad B_{2}(a_{1})=\left\{ 
\begin{array}{ll}
[0,1]\smallskip  & \text{if }a_{1}=0 \\ 
\{1\} & \text{if }a_{1}>0%
\end{array}%
\right. 
\end{equation*}%
Hence, the set of Nash equilibria of this game $\Gamma$ is given by $\left\{ (0,a_{2})\mid \tfrac{1}{2}\leqslant a_{2}\leqslant 1\,\right\} $ and the set of social optima is determined as $\{(1,a_{2})\mid 0\leqslant a_{2}\leqslant 1\,\}$. 
\\
Finally, using the formulations given in Example \ref{ex:First}, $\bar{v}^{\lambda }(\{1\})=0$, $\bar{v}^{\lambda }(\{2\})=\tfrac{1}{2}$, and $\bar{v}^{\lambda }(\{1,2\})=\max_{(a_{1},a_{2})\in [0,1]^{2}}\tfrac{a_{1}}{2} =\tfrac{1}{2}$. Therefore, $\mathcal{C}(\bar{v}^{\lambda })=\left\{ \,\left( 0,\tfrac{1}{2}\right) \,\right\} \neq \varnothing$, which selects from the set of Nash equilibria, but \emph{not} from the set of social optima in this game.
\end{example}

\singlespace

\subsubsection*{Statements:}

\paragraph{Acknowledgements}

We thank Emiliya Lazarova for developing with us the concepts on which this work is founded. This research has been funded under the Research Funding Program of University (FRA) 2022 of the University of Naples Federico II (GOAPT project) with the contribution of the Compagnia di San Paolo. Lina Mallozzi is member of the National Group for Mathematical Analysis, Probability and their Applications (GNAMPA) of the Istituto Nazionale di Alta Matematica (INdAM).

\paragraph{Declaration on conflict of interest}

The corresponding authors state that there is no conflict of interest.

\bibliographystyle{ecta}
\bibliography{RPDB}

\begin{thebibliography}{23}
\newcommand{\enquote}[1]{``#1''}
\expandafter\ifx\csname natexlab\endcsname\relax\def\natexlab#1{#1}\fi

\bibitem[\protect\citeauthoryear{Akerlof}{Akerlof}{1997}]{Akerlof1997}
\textsc{Akerlof, G.~A.} (1997): \enquote{Social distance and social decisions,}
  \emph{Econometrica}, 65, 1005--1027.

\bibitem[\protect\citeauthoryear{Aumann}{Aumann}{1959}]{Aumann1959}
\textsc{Aumann, R.~J.} (1959): \enquote{Acceptable {P}oints in {G}eneral
  {C}ooperative $n$-{P}erson {G}ames,} in \emph{Contributions to the {T}heory
  of {G}ames}, ed. by A.~W. Tucker and R.~D. Luce, Princeton University Press,
  vol.~IV, 287--324.

\bibitem[\protect\citeauthoryear{Balder}{Balder}{1997}]{Balder1997}
\textsc{Balder, E.} (1997): \enquote{Remarks on Nash equilibria for games with
  additively coupled payoffs,} \emph{Economic {T}heory}, 9, 161--167.

\bibitem[\protect\citeauthoryear{Chander}{Chander}{2010}]{Chander2010}
\textsc{Chander, P.} (2010): \enquote{Cores of games with positive
  externalities,} Discussion Paper 2010004, CORE, Louvain-la-Neuve, Belgium.

\bibitem[\protect\citeauthoryear{Chander and Tulkens}{Chander and
  Tulkens}{1997}]{Chander1997}
\textsc{Chander, P. and H.~Tulkens} (1997): \enquote{The core of an economy
  with multilateral environmental externalities,} \emph{International {J}ournal
  of {G}ame {T}heory}, 26, 379--401.

\bibitem[\protect\citeauthoryear{Chinchuluun, Pardalos, Migdalas, and
  Pitsoulis}{Chinchuluun et~al.}{2008}]{Chinchuluun2008}
\textsc{Chinchuluun, A., P.~M. Pardalos, A.~Migdalas, and L.~Pitsoulis} (2008):
  \emph{Pareto Optimality, Game Theory and Equilibria}, no.~17 in Springer
  Optimization and Its Applications, New York, NY: Springer Verlag.

\bibitem[\protect\citeauthoryear{Currarini and Marini}{Currarini and
  Marini}{2003}]{Currarini2003}
\textsc{Currarini, S. and M.~Marini} (2003): \enquote{A sequential approach to
  the characteristic function and the core in games with externalities,} in
  \emph{Advances in Economic Design}, ed. by M.~Sertel and S.~Koray,
  Heidelberg, Germany: Springer Verlag.

\bibitem[\protect\citeauthoryear{Currarini and Marini}{Currarini and
  Marini}{2015}]{Currarini2015}
---\hspace{-.1pt}---\hspace{-.1pt}--- (2015): \enquote{Coalitional approaches
  to collusive agreements in oligopoly games,} \emph{Manchester School}, 83,
  253--287.

\bibitem[\protect\citeauthoryear{Gillies}{Gillies}{1959}]{Gillies1959}
\textsc{Gillies, D.~B.} (1959): \enquote{Solutions to general non-zero-sum
  games,} \emph{Contributions to the Theory of Games}, 4, 47--85.

\bibitem[\protect\citeauthoryear{Hart and Kurz}{Hart and
  Kurz}{1983}]{HartKurz1983}
\textsc{Hart, S. and M.~Kurz} (1983): \enquote{Endogenous formation of
  coalitions,} \emph{Econometrica}, 51, 1047--1064.

\bibitem[\protect\citeauthoryear{Helm}{Helm}{2001}]{Helm2001}
\textsc{Helm, C.} (2001): \enquote{On the existence of a cooperative solution
  for a coalitional game with externalities,} \emph{International {J}ournal of
  {G}ame {T}heory}, 30, 141--146.

\bibitem[\protect\citeauthoryear{Lardon}{Lardon}{2012}]{Lardon2012}
\textsc{Lardon, A.} (2012): \enquote{The $\gamma$-core in Cournot oligopoly
  TU-games with capacity constraints,} \emph{Theory and Decision}, 72,
  387--411.

\bibitem[\protect\citeauthoryear{Lardon}{Lardon}{2019}]{Lardon2019}
---\hspace{-.1pt}---\hspace{-.1pt}--- (2019): \enquote{On the coalitional
  stability of monopoly power in differentiated Bertrand and Cournot
  oligopolies,} \emph{Theory and Decision}, 87, 421--449.

\bibitem[\protect\citeauthoryear{Lardon}{Lardon}{2020}]{Lardon2020}
---\hspace{-.1pt}---\hspace{-.1pt}--- (2020): \enquote{The core in Bertrand
  oligopoly TU-games with transferable technologies,} \emph{The BE Journal of
  Theoretical Economics}, 20, 1--10.

\bibitem[\protect\citeauthoryear{Le~Breton and Weber}{Le~Breton and
  Weber}{2011}]{LeBretonWeber2011}
\textsc{Le~Breton, M. and S.~Weber} (2011): \enquote{Game of Social
  Interactions with Global and Local Externalities,} \emph{Economics
  {L}etters}, 111, 88--90.

\bibitem[\protect\citeauthoryear{Milchtaich}{Milchtaich}{2009}]{Milchtaich2009}
\textsc{Milchtaich, I.} (2009): \enquote{Weighted congestion games with
  separable preferences,} \emph{Games and {E}conomic {B}ehavior}, 67, 750--757.

\bibitem[\protect\citeauthoryear{Mishra, Nath, and Roy}{Mishra
  et~al.}{2018}]{Mishra2018}
\textsc{Mishra, D., S.~Nath, and S.~Roy} (2018): \enquote{Separability and
  decomposition in mechanism design with transfers,} \emph{Games and {E}conomic
  {B}ehavior}, 109, 240--261.

\bibitem[\protect\citeauthoryear{Nash}{Nash}{1950}]{Nash1950}
\textsc{Nash, J.} (1950): \enquote{Equilibrium Points in $n$-Person Games,}
  \emph{Proceedings of the {N}ational {A}cademy of {S}ciences}, 36, 48--49.

\bibitem[\protect\citeauthoryear{Peleg}{Peleg}{1998}]{Peleg1998}
\textsc{Peleg, B.} (1998): \enquote{Almost all equilibria in dominant
  strategies are coalition-proof,} \emph{Economics {L}etters}, 60, 157--162.

\bibitem[\protect\citeauthoryear{Reddy and Zaccour}{Reddy and
  Zaccour}{2016}]{ReddyZaccour2016}
\textsc{Reddy, P.~V. and G.~Zaccour} (2016): \enquote{A friendly computable
  characteristic function,} \emph{Mathematical {S}ocial {S}ciences}, 82,
  18--25.

\bibitem[\protect\citeauthoryear{Stamatopoulos}{Stamatopoulos}{2016}]{Stamatopoulos2016}
\textsc{Stamatopoulos, G.} (2016): \enquote{The core of aggregative cooperative
  games with externalities,} \emph{The BE Journal of Theoretical Economics},
  16, 389--410.

\bibitem[\protect\citeauthoryear{Stamatopoulos}{Stamatopoulos}{2020}]{Stamatopoulos2020}
---\hspace{-.1pt}---\hspace{-.1pt}--- (2020): \enquote{On the $\gamma$-core of
  asymmetric aggregative games,} \emph{Theory and Decision}, 88, 493--504.

\bibitem[\protect\citeauthoryear{Zhao}{Zhao}{1999}]{Zhao1999Beta}
\textsc{Zhao, J.} (1999): \enquote{A $\beta$-core existence result and its
  application to oligopoly markets,} \emph{Games and {E}conomic {B}ehavior},
  27, 153--168.

\end{thebibliography}

\end{document}